\documentclass[letterpaper,twocolumn,10pt]{article}
\usepackage{usenix}

\usepackage{tikz}
\usepackage{nicematrix}
\usepackage{amsmath}
\usepackage{epsfig,endnotes}
\usepackage{xurl}      
\usepackage{hyperref}  
\usepackage{graphicx}
\usepackage{multirow}
\usepackage{booktabs}     
\usepackage{tabularx}     
\usepackage{adjustbox}    
\usepackage{lscape}       
\usepackage{makecell}
\usepackage{enumitem}
\usepackage[most]{tcolorbox}
\usepackage{graphicx}
\usepackage{threeparttable}
\usepackage{pgfplots}
\pgfplotsset{compat=1.17}
\usepackage{caption}
\usepackage{array}
\newcolumntype{C}[1]{>{\centering\arraybackslash}m{#1}}
\usetikzlibrary{positioning,arrows.meta}
\usepackage[table]{xcolor}
\newtcolorbox{tightbox}{
  colback=gray!10,    
  colframe=black,     
  boxrule=0.3pt,       
  arc=1pt,            
  left=2pt,right=2pt,   
  top=2pt,bottom=2pt
}

\usepackage[normalem]{ulem}

\usepackage{filecontents}

\begin{document}
\pagestyle{empty}

\date{}

\title{\Large \bf ``Abuse Risks are Often Inherent to Product Features'': \\ Exploring AI Vendors' Bug Bounty and Responsible Disclosure Policies}

\author{
{\rm Yangheran Piao}\\
University of Edinburgh\\
{\rm\small\texttt{lawrence.piao@ed.ac.uk}}
\and
{\rm Jingjie Li}\\
University of Edinburgh\\
{\rm\small\texttt{jingjie.li@ed.ac.uk}}
\and
{\rm Daniel W. Woods}\\
University of Edinburgh\\
{\rm\small\texttt{daniel.woods@ed.ac.uk}}
} 

\maketitle

\begin{abstract}
As vendors adopt AI technologies, security researchers are working to uncover and fix related vulnerabilities, which is important given AI systems handle sensitive data and critical functions. 
This process relies on vendors receiving and rewarding AI vulnerability reports.
To assess current practices, we analyzed the vulnerability disclosure policies of 264 AI vendors.
We employed a mixed-methods approach, combining snapshot and longitudinal qualitative analysis, as well as comparing alignment with 320 AI incidents and 260 academic articles. 
Our analysis reveals that 36\% of AI vendors have no established policy, and only 18\% mention AI risks.
Data access, authorization, and model extraction vulnerabilities are most consistently declared in-scope.
Jailbreaking and hallucination are most commonly declared out-of-scope. 
We identify three profiles that reflect vendors' different positions toward AI vulnerabilities: proactive clarification (\textit{n} = 46), silent (\textit{n} = 115), and restrictive (\textit{n} = 103). Our alignment results suggest that vendors may address AI vulnerability disclosure later than academic research and real-world incidents.

\end{abstract}

\section{Introduction}
Advances in machine learning (ML), large language models (LLM), and related techniques have been widely adopted in industry.
As a result, artificial intelligence (AI) is now deeply embedded in production software and critical systems.
Adoption has been accelerated by vendors making misleading claims about the capabilities of AI models~\cite{narayanan2024aisnake}.
The security implications of this transformation are still unclear~\cite{papernot2018sok}.

Researchers have published countless attacks on AI systems~\cite{tramer2016stealing,pang2022automl,hu2021ai,song2020embedding, yao2024llm}. 
It is an open question whether AI vulnerabilities are distinct from traditional software flaws, thereby motivating new disclosure policies~\cite{Cattell24,Householder24}. Unlike traditional systems composed of deterministic instructions explicitly coded by humans, AI systems operate on probabilistic rules learned from vast quantities of data~\cite{bender2021dangers,bommasani2022opportunitiesrisksfoundationmodels}. AI vulnerabilities might be special because they exploit rules inferred from data and generative capabilities that sit outside traditional threat models~\cite{papernot2018sok,barreno2006can,biggio2018wild,wei2023jailbroken,hui2024pleak}. Alternatively, an outcome-oriented perspective says the mechanism of failure matters less than whether security properties are violated~\cite{schneier2015secrets,anderson2010security}. A command that leaks sensitive data constitutes a vulnerability regardless of whether it exploits an SQL or an LLM injection. In this view, AI vulnerabilities are not special.

These theoretical questions become real when researchers submit AI vulnerabilities to vendors~\cite{grotto21,Householder24}.
There are open questions about whether vendors will pay for prompt injections, whether models can be “patched”, and whether content safety issues like models generating hallucinations and abuse are in scope~\cite{narayanan2024aisnake,hiroshima23,whai23,carlini2021extracting,tramer2020adaptive}.
If AI vulnerabilities are special, vulnerability disclosure policies should specifically address AI security~\cite{Cattell24,Householder24,Walshe23,ahmed21}.
Alternatively, AI vendors may not feel the need to update disclosure policies specifically for AI vulnerabilities, as long as such vulnerabilities can be framed as violations of traditional security goals, such as enabling RCE in production systems.~\cite{Boucher22,de23,chen25}.
Vendors need to have strong security processes to manage the transition to AI security.
Prior work has found security practices around AI remain fragmented and poorly understood~\cite{Apruzzese23}. 
Vendors may also differ widely in whether, and how, they recognize such vulnerabilities~\cite{longpre2025inhouse,wang2025voluntary}.
For example, major vendors were slow to act on an NLP vulnerability report~\cite{Boucher22}.

In traditional software security, vulnerability reporting has been facilitated by bug bounty programs (BBPs) and vulnerability disclosure programs (VDPs).
These mechanisms have proven effective in incentivizing researchers and helping vendors improve security~\cite{luna2019productivity,walshe2020empirical}.
Prior empirical studies examined BBPs' and VDPs' disclosure policies before the rapid adoption of AI techniques~\cite{zhao18,Walshe23}.
It is unclear whether these mechanisms can translate effectively to AI-specific vulnerabilities~\cite{Cattell24,Hall25}.

To address this gap, our study analyzes the vulnerability disclosure policies of 264 AI vendors. We conducted snapshot and longitudinal measurements, and qualitatively analyzed the policies of major AI vendors.
We ask the following research questions and find the following:

\begin{itemize} \itemsep0em
    \item[\textbf{RQ1}] \textit{What is the state of vulnerability disclosure in the AI industry, and how has it evolved over time?}\\
    36\% of AI vendors have no disclosure channel. Only 18\% of vendors explicitly address AI security in disclosure documents, with the first mention in 2018. Vulnerabilities defined in terms of security goals are most consistently in scope (91\%). Vulnerabilities in AI models are less consistently in-scope, such as prompt injection (71\%), adversarial examples (80\%) and model extraction (90\%). Finally, jailbreaking (27\%), harmful output (45\%) and hallucination (17\%) are rarely in scope.
    \item[\textbf{RQ2}] \textit{How do vendors approach AI vulnerabilities?}\\
    Vendors can be classified into: (i) \textit{proactive} vendors (17\%), typically model providers, that clarify which AI reports are in/out of scope; (ii) \textit{silent} vendors (44\%), typically Gen AI apps, that do not mention AI reports; (iii) \textit{restrictive} vendors (39\%), often data science providers that reject AI reports or lack disclosure channel.
    \item[\textbf{RQ3}] \textit{What is the alignment with AI incidents and research?} \\Vendors and academics tend to address technical vulnerabilities, while most public incidents concern content safety. Vendors clarified AI vulnerability disclosure later relative to academia and real-world incidents.
\end{itemize}

\begin{figure*}[]
\hspace*{-0.3cm}
    \centering
    \includegraphics[width=0.9\textwidth]{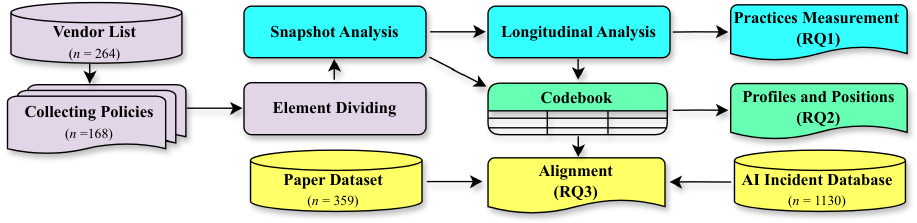}
    \caption{Methodology overview.}
    \label{fig:method}
\end{figure*}

\section{Related Work}
\label{sec:Background}
Software vendors have established mechanisms to encourage bug submissions.
VDPs can offer recognition and some degree of safe harbor from legal prosecution for researchers who report vulnerabilities~\cite{eurosp24,USENIX16,piao2025cfaa}.
BBPs additionally offer financial incentives for such reports~\cite{zhao17,ahmed21}.
These initiatives not only enhance the overall security posture of vendors but also facilitate more structured and responsible processes for handling vulnerabilities~\cite{Liu25,ding19,sridhar2021hacking}. BBPs increase the volume of reported vulnerabilities~\cite{luna2019productivity,zhao2015empirical}, often at a lower cost than internal software security investments~\cite{walshe2020empirical,walshe2022coordinated}. 

\textbf{Communication Challenges.}
Research into the perspectives of bug hunters has revealed communication challenges~\cite{hata2017understanding,votipka2018hackers,albanna2018friendly}. 
Problems like delayed responses, disputes over severity, and limited feedback are consistently identified as sources of frustration~\cite{akgul2023bug,piao2025unfairness,piao2025study}.
Similarly, academic researchers have experienced inconsistent disclosure processes and even hostility~\cite{sriramulu2025poster}.
Even when documents are found, the terms and conditions are often ambiguous~\cite{fulton2023vulnerability,piao2025unfairness}.

\textbf{Policy Analysis.}
Laszka \textit{et al}.~\cite{zhao18} conducted the first systematic study of BBP policies, highlighting the importance of rule transparency and readability for program effectiveness.
However, the sample, collected in 2016, was limited to one platform, HackerOne. 
Walshe \textit{et al.}~\cite{Walshe23} expanded their study to VDPs, which were classified into a 12-element framework. 
The corpus~\cite{Walshe23} was collected before most vendors had clarified BBPs and VDPs to address AI vulnerabilities, as our results will show (see Figure~\ref{fig:evolution}).

\textbf{AI Vulnerabilities}
The rapid integration of AI into products and services has created security risks~\cite{papernot2018sok,pang2022automl,hu2021ai,song2020embedding}.
This has given rise to new attacks like: adversarial examples~\cite{yuan2019adversarial,biggio2018wild,yao2024llm}; backdoors~\cite{naseri2024badvfl,lin2020composite,jia2022badencoder}; inference attacks~\cite{rigaki2023privacy,luo2022shapley}; prompt injection~\cite{shi2024promptinjection,das2025llm}; data poisoning~\cite{tian2022poisoning,cong2024testtime}; jailbreaking~\cite{yang2024sneakyprompt,deng2023masterkey}; model inversion and extraction~\cite{zhao2024loki,dibbo2023sok,oliynyk2023stealing,carlini2021extracting,an2022mirror,hui2024pleak}; and content safety and bias~\cite{tang2024gendercare,si2022toxic}.
AI vulnerabilities often exploit probabilistic systems and sit outside traditional threat models, which complicate their classification, reporting, and even remediation~\cite{Householder24}. 

\textbf{AI Disclosure Challenges.} Much like with disclosure of traditional vulnerabilities~\cite{sriramulu2025poster}, Boucher \textit{et al.}~\cite{Boucher22} received little response to their report of various commercial NLP models being vulnerable to ``a single imperceptible encoding injection''. Most model providers did not implement fixes at that time~\cite{Boucher22}.
AI vendors have been criticized for lagging in adopting AI security measures~\cite{wang2025voluntary}, relying on ``security through obscurity''~\cite{Hall25}, and the lack of safe harbors for security research~\cite{SafeHarbor24}.
This has motivated calls for standardized reporting, legal protections, and supply-chain relays~\cite{longpre2025inhouse}, covering issues like bias, validity, and misuse~\cite{Cattell24}.

\textbf{Research Gap.}
There has been no systematic study of how disclosure policies treat AI vulnerabilities, and no general study of policies since 2022~\cite{Walshe23}.
To address this gap, our work examines how AI companies adapt bug bounty rules and disclosure policies to address emerging AI vulnerabilities, and how this has evolved over time.
This is motivated by recent research interest in AI security~\cite{papernot2018sok,pang2022automl,hu2021ai,song2020embedding}, and arguments that AI vendors' security policies are inadequate~\cite{wang2025voluntary, Hall25, SafeHarbor24}.
Anecdotal evidence can be seen in the challenges a research team faced in disclosing AI issues to major model providers~\cite{Boucher22}.

\section{Methodology}
We studied how AI vendors address vulnerability disclosure using multiple data sources (see Figure~\ref{fig:method}). We define \textit{AI vendors} to be companies that build products with substantial AI capabilities\footnote{Our definition includes providers of frontier models, GenAI applications, and AI/ML training infrastructure. It does not include companies that merely install AI chatbots on their websites; or sell AI consulting.}.
To answer \textbf{RQ1} and \textbf{RQ2}, we collect a corpus of disclosure policies published by AI vendors. 
Section~\ref{sec:Collection} describes how we selected vendors and sampled documents.
Our analysis is described in Section~\ref{sec:Assigning}.
To answer \textbf{RQ3}, we conduct a secondary analysis of a database of AI incidents and a list of academic articles, which is described in Section~\ref{rq3}.

\subsection{Data Collection}
\label{sec:Collection}
\textbf{Identifying AI vendors.} Many vendors incorporate ``AI'' into their branding and marketing, even when their offerings use minimal AI techniques~\cite{narayanan2024aisnake}. 
Relying exclusively on such self-reported claims would risk including vendors that are not substantively engaged in the development or deployment of artificial intelligence. 

We first collected product lists across nine AI-related categories (described in Appendix~\ref{sec:Supplement}) published by Gartner~\cite{GartnerPeerInsights}, an internationally recognized authority in technology market analysis.
These lists are subject to continual review, which allows companies to challenge their exclusion or request reevaluation. Gartner has a commercial incentive to maintain comprehensive and accurate coverage of AI products, given their customers use these lists for procurement decisions.

Some vendors offer multiple AI products, so we performed merging and deduplication to ensure each company was represented only once in our dataset. 
This resulted in a list of 264 unique companies\footnote{\url{https://doi.org/10.5281/zenodo.17873559}}, which serves as the basis for our analysis.
To assign categories to vendors, we used a priority-based mapping where ``core'' categories were assigned first.
For example, vendors who offer foundation models are ``model providers'', even if they also offer an AI code assistant. 
This ensures the classification reflects the most significant products provided by each company, reducing noise from overlapping offerings.

\textbf{Collecting Disclosure Policies.}
Most vulnerability disclosure policy documents are published as web pages.
To identify whether a mechanism for third-party vulnerability reporting exists, we searched for the disclosure policy across: a) \textit{security.txt} files\footnote{Security.txt is a standard that helps websites share security policies and receive vulnerability reports.} from the Firebounty platform; b) examining major bug bounty platforms, including HackerOne, Bugcrowd, and Huntr; and c) manually reviewing web search engine results.
We then cross-checked each company's official website to minimize the risk of missing relevant information. 

We found that 96 companies in the sample lacked any kind of vulnerability disclosure mechanism. Two of these vendors may have shut down because their website was not accessible. For companies without any public vulnerability disclosure channel, we additionally examined their security-related webpages to determine whether they included any relevant statements or implied mechanisms for receiving reports.

For the AI companies that allow vulnerability submissions, we collected all available vulnerability disclosure and bug bounty policies through a semi-automated process. For websites that prohibit automated crawling, we used manual collection. Five companies published multiple disclosure documents across different platforms, and we merged their documents according to the corresponding structure. 

In total, we gathered 168 vulnerability-related policy documents. We then classified the documents based on: 1) classifying a document as a BBP if it offers financial rewards; 2) classifying it as a VDP if it describes the scope and rules around submitting security vulnerabilities; 3) classifying it as a security contact if it was just a contact email or channel with a few lines of text (e.g. in a security.txt file).

\subsection{Disclosure Policy Analysis (RQ1, RQ2)} 
\label{sec:Assigning}
We conducted a two-level analysis to avoid analyzing AI-specific content without considering the overall structure of the document.
Analyzing both levels accounts for the modular nature of disclosure policies, while also capturing granular details about how AI vulnerabilities are treated.
Section~\ref{sec:3.2.1} describes how we mapped out the high-level structure using a taxonomy from prior work~\cite{Walshe23}.
Section~\ref{sec:AIMention} explains how we analyzed AI-specific aspects of each document.

\subsubsection{High-Level Structure} 
\label{sec:3.2.1}
Disclosure policies typically use similar headings and sub-headings (e.g. scope, eligibility, and engagement rules). This structural regularity made it possible to systematically align each document with established policy elements, an approach used in prior work~\cite{zhao18,Walshe23}. We rely on these structural cues (headings, sub-headings, and section organization) when mapping the documents to analytical elements.

We organized the 168 documents using Walshe \textit{et al.}~\cite{Walshe23}'s analytical framework that categorizes bug bounty policies into 12 elements. 
These include scope, eligibility of vulnerabilities, rules of engagement, legal \& procedural guidelines, restrictions on participants \& activities, evaluation of rewards, and official company statements. To identify structural themes from each document, we manually mapped (sub)headings to the prior work's codebook~\cite{Walshe23} (introduced in Appendix~\ref{sec:Supplement}).

After reading the policies in full, we found that the previous framework required extensions to better address our research questions.
During team discussions among three researchers, we identified several policy components that repeatedly appeared as standalone structural sections across documents but were not explicitly and fully captured in the prior framework. 

As a result, we added \verb|Service-Commitments| to capture whether vendors specify different response timelines for AI-related reports, and \verb|Contact-Channel| to reflect the practice of directing AI vulnerability submissions to separate reporting channels. In addition, we split \verb|Legal-Clauses| into: \verb|Compliance-Requirements| and \verb|Safe-Harbor|, as these two aspects serve distinct functions: safe harbor provisions are typically separated from general legal clauses in policies to address the special considerations associated with protecting security researchers~\cite{Hantke24}. 
Adding these elements allowed us to identify differences in how vendors structure AI-specific disclosure practices.

All collected policy texts were mapped according to our expanded 15-element schema. To account for modularity, we segmented the text using subheadings. 
For policies that did not contain a particular element, the corresponding field was left empty. This process resulted in 1,011 policy elements. We did not test IRR for the existing structural elements as the codebook was already established in prior work~\cite{Walshe23}.

For the four new elements, we ensured the reliability through team discussions rather than independent parallel coding. Coding these extended elements required minimal interpretation because they were identified based on explicit structural subheadings, which were consistent across the documents. For example, AI reporting channels are often listed separately from standard submission instructions. We manually reviewed all documents to validate the accuracy of elements assignments and discussed ambiguities as a research team, which were infrequent given the clarity and standardized structure of disclosure policies.

\subsubsection{AI-specific Analysis} 
\label{sec:AIMention}
During manual review, we identified 48 vendors that explicitly mentioned AI-related vulnerabilities in their policies. The policies of these 48 AI companies form the sample used in our AI-specific qualitative analysis.

\textbf{Snapshot Analysis.}
To analyze the most recent policies, we conducted keyword searches to extract all text segments related to AI vulnerabilities or AI products from policies. 
To identify terms, we relied on glossaries published by authoritative sources which are provided in Appendix~\ref{sec:Supplement}. We manually reviewed the text surrounding keywords to ensure that complete and meaningful excerpts were collected, and used the extended excerpts as units of analysis during thematic coding. After reviewing, we organized and structured the text into discrete analytical units, each representing a complete and coherent thematic idea, resulting in a total of 422 units.
We grouped these units by both vendor type and the corresponding element they addressed. 

We developed a codebook based on an initial review and collaborative discussion among the research team. 
To ensure the reliability and consistency of our coding framework, two authors independently applied the codebook to a randomly selected sample of 100 analytical units. 
The coded responses achieved a high Cohen's Kappa (\textit{k} = 0.816), indicating strong agreement between coders. 
The discrepancies in coding were clarified and discussed by the research team.

Following refinement, the first author proceeded to code the remaining analytical units using the final version of the codebook. This version was used as the basis for the themes presented in the results section. The codebook is provided in the Appendix~\ref{sec:codebook}.

\textbf{Longitudinal Analysis.}
For the 48 sampled AI vendors, we also collected historical policy data to track changes over time. We retrieved archived versions of their vulnerability submission policies from the Internet Archive and the changelog provided by bounty platforms. 
We used a Python script based on the WaybackPy library\footnote{\url{https://pypi.org/project/waybackpy/}} to automatically identify sections of policy text that had changed across historical versions. 
For two vendors whose archived pages were not available, we collected public announcements related to their bug bounty programs to identify the approximate time when AI-related vulnerabilities were first incorporated, as well as subsequent modifications and the specific changes made. 

Finally, we examined each policy update and extracted a total of 205 distinct changes involving AI-related policy elements.
We applied the same AI keyword procedure used in the snapshot analysis to determine which changes were AI-related.
All extracted change units were analyzed using the same codebook developed for the snapshot analysis. No new themes emerged during this phase, and ambiguities were resolved through group discussion among three researchers.

\subsection{Incidents and Research Trends (RQ3)}
\label{rq3}

Our final research question asks whether AI security issues in disclosure policies are aligned with actual incidents and academic research.
Comprehensively collecting and analyzing primary data about either would represent a standalone contribution outside the scope of this paper.
Instead, we conducted a secondary analysis of an AI incident repository~\cite{AAAI21}, and a database of AI security articles published at the four major security conferences\footnote{\url{https://github.com/gnipping/Awesome-ML-SP-Papers}\label{fn:big4}}.

\begin{figure}[t]
    \centering
    \includegraphics[width=0.45\textwidth]{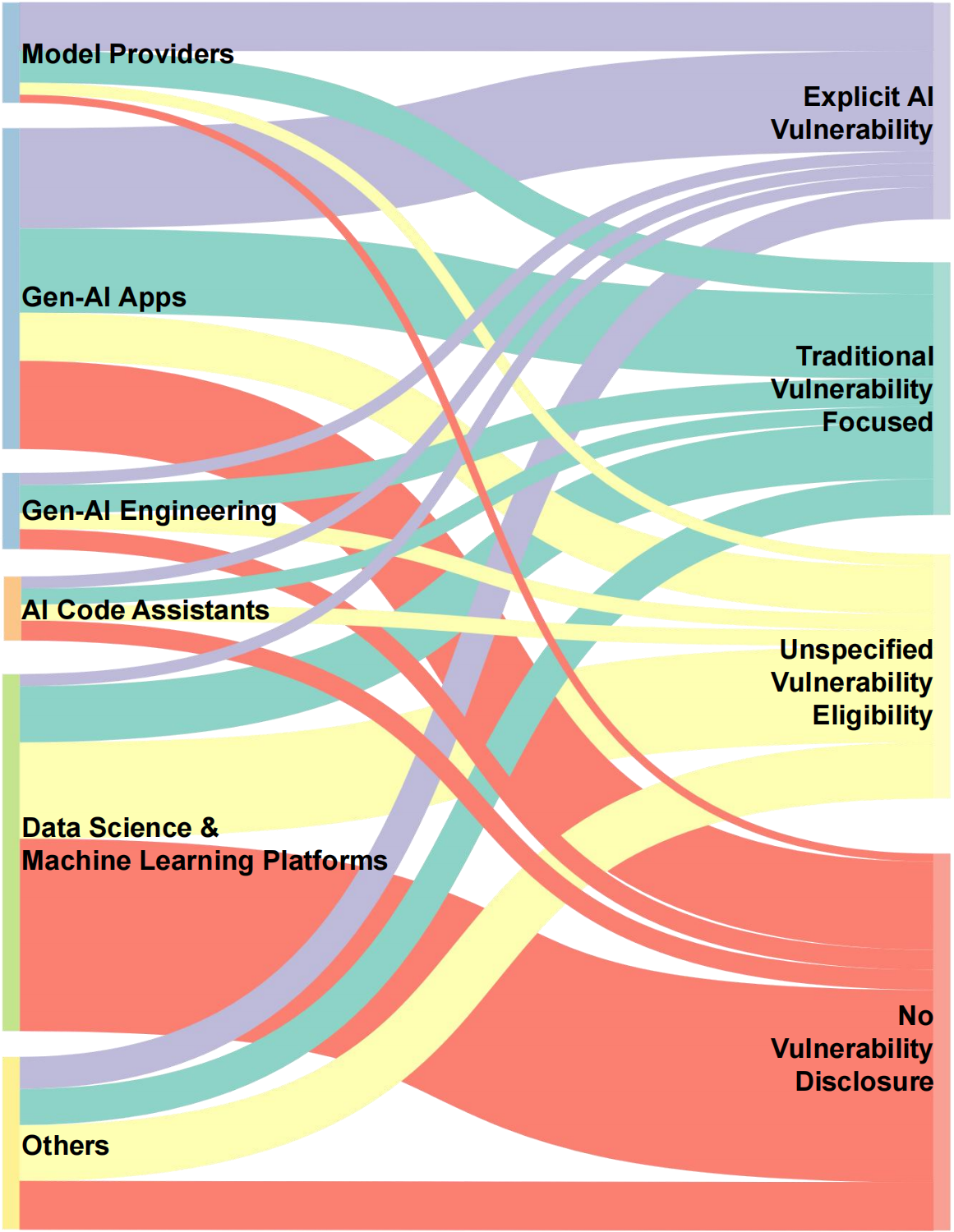}
    \caption{Mapping AI company categories to vulnerability disclosure approaches.}
    \label{fig:size}
\end{figure}

\textbf{Incidents.} We analyzed data from the AI Incident Database (AIID)~\cite{AAAI21}, a public, community-driven repository of AI failures. 
The database collects incidents submitted by journalists, users, and researchers.
The corpus captures a broader notion of security than the CIA triad, covering various harms resulting from AI systems like discrimination, misinformation, and socioeconomic \& environmental harms.
Each incident is manually reviewed and enriched with metadata, including tags and categorical labels from different taxonomies. We extracted 1,130 incidents from the database up to and including December 2024.

To structure the analysis, we adopted the AIID's MIT AI Risk Taxonomy~\cite{Risk24}, a comprehensive framework introduced to systematize risks from AI across multiple dimensions. We also mapped the AI vulnerability types mentioned in the policies to the MIT taxonomy, with the method described in Appendix~\ref{sec:MIT}.

After deduplicating the AIID entries and mapping the incidents to taxonomy labels, we linked them to 47 AI vendors identified in the previous section, resulting in a subset of 320 relevant incidents. We compared these filtered real-world AI incidents against our codebook to assess alignment between reported events and AI vulnerability disclosure policies.

\textbf{Academic Research.} 
We wanted to understand which AI issues the research community is focused on.
We conducted a comprehensive literature review of academic taxonomies and surveys on AI security to extract a meta-taxonomy of AI vulnerabilities. The literature search resulted in 146 articles from academic databases. 
We applied inclusion criteria focused on offensive security and broad classification, manually screening for relevance. 
We excluded research focused on detection, defense, or usability. Details on the collection criteria and the resulting meta-taxonomy can be found in Appendix~\ref{sec:Taxonomy}.

To verify the completeness of the resulting meta-taxonomy, we analyzed 359 AI security and safety papers from a dataset of four security conferences (2020–2024)\footref{fn:big4}.
Finally, this resulted in a six-category meta-taxonomy of AI vulnerabilities, as well as the year-by-year thematic distribution of AI security papers over the past four years based on this taxonomy.

\section{RQ1: Reporting Practices Measurement}
\label{sec:4}
This section describes how disclosure documents address AI security. Section~\ref{sec:Status} focuses on whether, and through which channel, AI issues can be disclosed to a vendor.
Section~\ref{sec:PSD} presents the structure of disclosure policies, with a focus on the characteristics and pricing of policies that mention AI risks. 
Section~\ref{sec:vuln} explores granular questions like which specific vulnerabilities and risks are in/out of scope.
Finally, Section~\ref{sec:Evolution} illustrates evolving trends in AI-related policy elements over time.

\subsection{Current Status}
\label{sec:Status}
\textbf{How to Disclose a Vulnerability.} The adoption of vulnerability disclosure mechanisms among AI companies (64\%) is lower than the adoption of BBPs among Fortune 100 vendors (72\%)~\cite{vmr23}. 
This is likely because many AI vendors are young companies who prioritize growth over security processes.

Across 264 AI vendors, 36\% are found to have no public vulnerability reporting or disclosure channel. 
Manual review revealed that a quarter of these vendors at least provided security statements or trust centers on their websites\footnote{The trust center is an emerging type of public-facing statement that outlines a company's internal security, privacy, and compliance practices.}.
Such vendors may rely on internal security research or contracted pentesting.
This still leaves more than a quarter (27\%) of AI vendors who have no reporting channel or public security information.
Around 16\% of AI vendors operate VDPs, which accept submissions without offering monetary rewards. VDPs at least provide researchers with some degree of protection from legal prosecution. Additionally, 21 vendors have not established a formal reporting process but provide a dedicated email address or channel for vulnerability submissions.

Finally, 40\% of AI vendors run a BBP that offers rewards for valid vulnerability reports.
Many of these vendors rely on a platform to host their public BBP. In contrast, 35 independently operate their own bug bounty programs, a practice more common among larger companies.
Additionally, 10 vendors operate private programs where the scope is not publicly visible and could only be accessed by invited researchers. 

\begin{table}[t]
\centering
\begin{threeparttable}
\captionsetup{skip=6pt}
\caption{Coverage of policy elements across AI vendors}
\label{tab:element_coverage}
\begin{tabular}{p{0.48\columnwidth}c >{\centering\arraybackslash}p{0.1\columnwidth} >{\centering\arraybackslash}p{0.1\columnwidth}}
\toprule
{\textbf{Policy Elements}} & {\textbf{Trad}} & {\textbf{AI} } & {\textbf{$\Delta$}} \\
\midrule
\verb|Contact-Channel|         & 100\% & 100\% & 0\% \\
\verb|Company-Statement|       & 92\% & 100\% & +8\% \\
\verb|Guideline-Submission|    & 76\% & 79\% & +3\% \\
\verb|Reward-Evaluation|       & 56\% & 69\% & +13\% \\
\verb|Scope-In|                & 49\% & 88\% & \cellcolor{gray!30}+39\% \\
\verb|Prohibited-Action|       & 51\% & 60\% & +9\% \\
\verb|Vuln-Ineligible|         & 41\% & 79\% & \cellcolor{gray!30}+38\% \\
\verb|Guideline-Disclosure|    & 50\% & 56\% & +6\% \\
\verb|Engagement|              & 39\% & 73\% & \cellcolor{gray!20}+34\% \\
\verb|Vuln-Eligible|           & 42\% & 56\% & +14\% \\
\verb|Scope-Out|               & 38\% & 50\% & +12\% \\
\texttt{Service-Commitment}      & 31\% & 65\% & \cellcolor{gray!20}+34\% \\
\texttt{Compliance-Requirements} & 36\% & 46\% & +10\% \\
\verb|Safe-Harbor|             & 18\% & 67\% & \cellcolor{gray!48}+48\% \\
\verb|Participant-Restriction| & 28\% & 38\% & +10\% \\
\bottomrule
\end{tabular}
\begin{tablenotes}
\footnotesize \item Trad: Proportion of elements in traditional policies (no AI mentioned). 
\footnotesize \item AI: Proportion of  elements in policies that mentioned AI. 
\footnotesize \item $\Delta$: The difference between same elements in Non-AI and AI.
\end{tablenotes}
\end{threeparttable}
\end{table}

\textbf{Disclosure Policies by Type of AI Vendor.} Figure~\ref{fig:size} shows that just 18\% of vendors in our sample explicitly mentioned AI issues. 
Model providers had the highest focus on AI vulnerabilities, and around half (48\%) explicitly mentioned AI vulnerabilities or included AI products within their defined scope. 
Only a small fraction of model providers (8\%) had no vulnerability disclosure mechanism. 

Many AI vendors (24\%) focused on traditional vulnerabilities without specifically mentioning AI security.
Generative AI Engineering vendors were the most likely to mention only traditional vulnerabilities, with 37\% doing so. 
The category most often lacked a vulnerability disclosure channel was Data Science \& Machine Learning Platforms, for which the majority (68\%) of vendors did not have a channel.
These kinds of products may be harder for attackers to access and exploit compared to an internet-accessible Gen-AI app or model.

The remaining AI vendors (22\%) could not be categorized because they did not clearly mention any type of vulnerability. 
These policies were VDPs with minimal description of scope.
They also included vendors that only provided a security contact email, lacking any formal disclosure policy. 

\subsection{Policy Structure}
\label{sec:PSD}

Turning to the high-level structure of documents, Table~\ref{tab:element_coverage} compares coverage of elements among all documents with coverage in policies that mention AI issues. 

\textbf{Completeness.}
All policies contained a designated channel for submitting reports (i.e. through the platform, online form or by contacting an email address), which is a necessary requirement for receiving bug reports.
Beyond that, the most frequent element was the \verb|Company-Statement|, by which vendors articulate a general commitment to security collaboration. 
However, many disclosure documents lacked elements.
Nearly 40\% of AI companies incorporated fewer than six elements, while only 6\% covered the full set of policy elements. 
Model providers had the most complete policies among all AI company types, with 42\% including more than ten elements and 25\% covering all elements comprehensively.

\textbf{Traditional vs AI-specific Policies.}
Intuitively, policies explicitly referencing AI vulnerabilities were more likely to contain all other elements.
For example, 88\% of AI policies included a clear \verb|Scope-In| definition that clarifies which systems or products are considered in scope. This is 39\% higher than the rate for policies that did not mention AI. Similarly, 79\% of AI-mention policies specified \verb|Vuln-Ineligible| conditions, which is higher than the rate of 38\% among traditional policies. 
Unsurprisingly, companies with more advanced disclosure documents were more likely to mention AI.

\textbf{Legal Gaps.} 
Legal or compliance-related elements were scarce. 
Only 36\% of traditional policies mentioned \texttt{Compliance-Requirement}. Among them, three quarters did not specify any concrete laws or regulations. Instead, they asked researchers to comply with broad obligations like ``\textit{all applicable laws and regulations,... governing privacy or the lawful handling of data,}'' which offers limited clarity to hackers. 
Only 9\% of traditional policies mentioned specific laws like the \textit{GDPR}, \textit{CFAA}, \textit{CCPA}, \textit{DMCA}, and \textit{COPPA}. 
Mentioning compliance requirements was more common (46\%) in AI-specific policies, with some mentioning the \textit{EU AI Act}. 

AI-specific policies were more than twice as likely to offer some form of \verb|Safe-Harbor| clause, compared to 18\% across traditional policies.
Only 28\% of traditional policies specified \verb|Participant-Restriction| clauses (vs. 38\% of AI-specific policies), often noting that individuals from certain countries or under specific legal conditions (e.g., export control restrictions) may be ineligible for participation or rewards.

\textbf{Interaction Protocols.} 
The absence of structured interaction protocols remained a concern. Only 31\% of traditional policies included a \verb|Service-Commitment| element, which typically outlines expected timelines for vendor responses or resolution. Response windows also varied widely, from fixed timelines like ``\textit{within 90 days}'' to vague phrasing such as ``\textit{we will get back to you as soon as possible.}''
The majority of AI-specific policies (65\%) provided a concrete timeline for responding to vulnerability reports.
Moreover, 73\% of AI-specific policies included \verb|Engagement| provisions (vs. 39\% of traditional policies). Some of these provided requirements for AI testing, such as: ``\textit{Any actions that could potentially harm our LLM systems... are strictly prohibited.}''

\textbf{Rewards.} 
The \verb|Reward-Evaluation| element captures pricing of vulnerabilities. 
After excluding private programs, programs that do not accept AI vulnerabilities, and those that do not offer monetary rewards, only 33 bug bounty programs offered publicly accessible reward information for AI vulnerabilities.
Among the 33 BBPs, 20 vendors applied a unified pricing structure, treating AI vulnerabilities the same as traditional ones in terms of severity tiers and reward brackets.

\begin{table}[t]
  \centering
  \caption{AI Vulnerability types with eligibility}
  \label{tab:vuln-types}
    \rowcolors{2}{gray!15}{white}
  \begin{tabular}{@{}ccc@{}}
    \toprule
    \textbf{Out-of-Scope} & \textbf{AI Vulnerability Type} & \textbf{In-Scope} \\
    \midrule
     1& System Integrity &5 \\
     1& Supply Chain &4 \\
     1& Resource Consumption &1 \\
     0& Data Access &9 \\
     0& Authorization &8 \\
     0& Authentication &2 \\
     \hline
     4& Prompt Injection &10 \\
     1& Model Extraction &9 \\
     8& Jailbreaking &3 \\
    1 & Inference & 4\\
     1& Data Poisoning &6 \\
     2& Data Leakage &4 \\
     1& Adversarial Example &4 \\
     \hline
     1& Policy Violations &4 \\
    2& Information Disclosure &7 \\
     12& Harmful/Insecure Output &10 \\
     6& Hallucination &1 \\
     4& Others &3 \\
    \bottomrule
  \end{tabular}
\end{table}

The remaining 13 vendors offered separate price ranges for AI vulnerabilities. 
Four vendors assigned lower payouts to AI-related issues, possibly reflecting lower perceived impact or greater ease of finding vulnerabilities. In contrast, nine companies assigned higher payouts to AI vulnerabilities than to most traditional vulnerabilities at comparable severity levels. These companies often classify AI products as ``\textit{core assets}''. However, the highest reward offered for AI issues did not exceed the top-tier bounty for the most severe traditional vulnerabilities. Even in cases where AI issues are priced higher than majority of traditional vulnerabilities, they still capped out at the same level as traditional ``\textit{Tier-1}'' assets. 

\subsection{AI-Specific Vulnerabilities}
\label{sec:vuln}
We classified the 140 AI-specific vulnerabilities extracted from policies into three major categories: AI systems, AI models, and AI features. Table~\ref{tab:vuln-types} displays the number of times these vulnerabilities are considered in or out of scope across vendor policies.

Among vendors that mentioned AI vulnerabilities, the largest share belonged to the model providers (60\%), followed by Gen-AI App (26\%) and Gen-AI Engineering (5\%). Model providers also accounted for the highest proportions of vulnerabilities addressing AI systems, AI models, and AI feature vulnerabilities, at approximately 79\%, 59\%, and 56\% respectively.

\textbf{AI System Vulnerabilities.} The vulnerabilities in this category do not explicitly name the type of AI attack.
Instead they are defined in terms of the security goals of an AI system. If the goal is violated, the bug is in scope. They were mentioned 32 times across policies, accounting for 23\% of all AI-specific vulnerabilities.
Qualifying vulnerabilities include those that violate
system-level integrity (\textit{n} = 6), data access (\textit{n} = 9), supply chain security (\textit{n} = 5), authentication and authorization mechanisms (\textit{n} = 10), and resource consumption (\textit{n} = 2). 
Attacks in this category often leverage conventional security flaws that take on new forms when applied to AI-powered services, such as ``\textit{focused on machine learning model file formats}'' stated by LangChain.

\textbf{AI Model Vulnerabilities.} Most vulnerabilities (41\% of AI-specific vulnerabilities) are defined in terms of model exploits. These terms are closer to those used in the academic community~\cite{papernot2018sok}.
They result from the unique characteristics of models like the inputs/outputs, and stage of the training/deployment pipeline.
Common vulnerabilities include prompt injection (\textit{n} = 14), model extraction (\textit{n} = 10), inference attacks (\textit{n} = 5), data poisoning (\textit{n} = 7), and jailbreaks (\textit{n} = 11). 
These model-level attacks aim to bypass safety constraints or steal proprietary assets.

\begin{figure*}[]
\hspace*{-0.3cm}
    \centering
    \includegraphics[width=\textwidth]{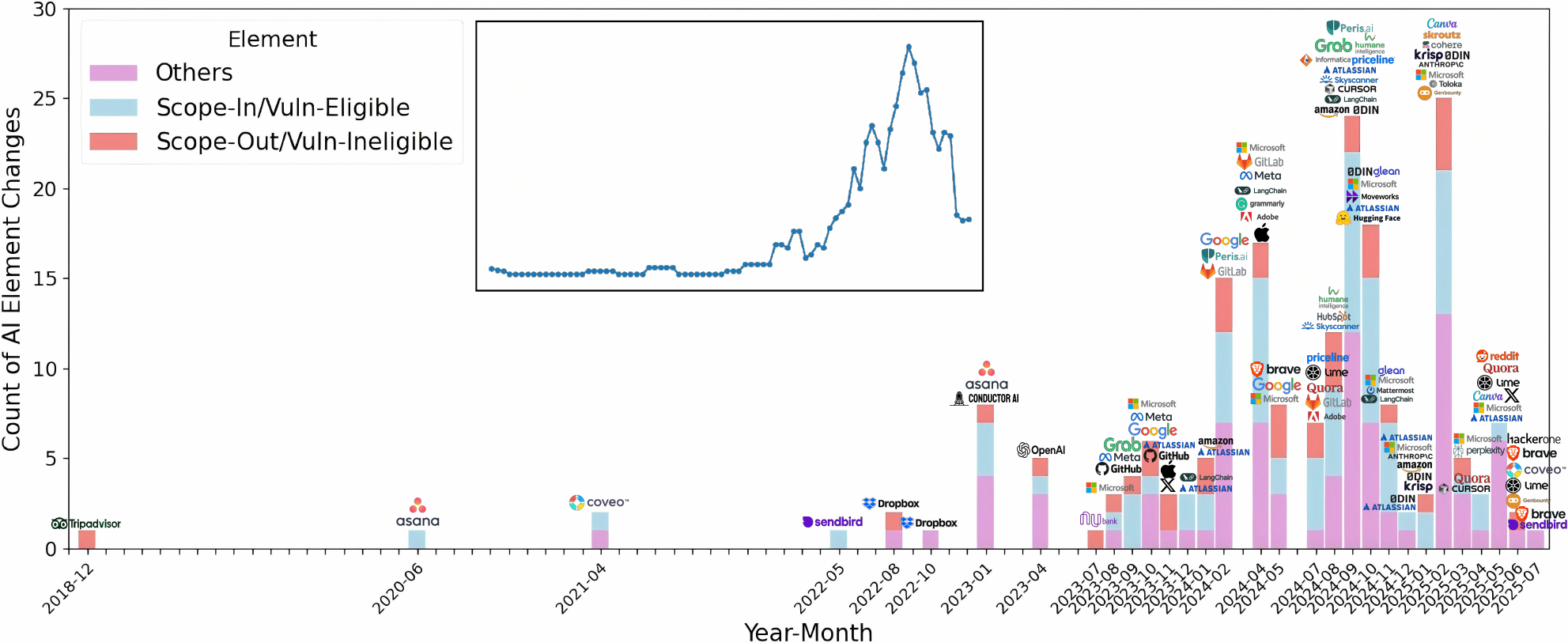}
    \caption{Monthly cumulative AI mention updates by policy element.}
    \label{fig:evolution}
\end{figure*}

\textbf{AI Features.} Of the remaining ``vulnerabilities'', they refer to functionality specific to generative AI systems, such as content generation, dialogue capabilities, or contextual personalization. Vulnerabilities in this category account for 36\% of all mentions of AI vulnerabilities. This includes harmful/insecure outputs (\textit{n} = 22), hallucinations (\textit{n} = 7), and information disclosure through unexpected behavior (\textit{n} = 9). 
These issues often exploit the interactive interfaces of models, leading the model to produce unsafe content, broadly defined.
For example, Amazon's policy describes content that is ``\textit{susceptible to solicitation and social engineering.}''
These are likely better described as safety problems than security vulnerabilities, a topic that we return to in Section~\ref{sec:discussion}.

\textbf{In vs Out of Scope.} 
We analyzed whether specific types of AI-related vulnerabilities were explicitly acknowledged as eligible within each vendor's disclosure policies. 
At a high level, AI system vulnerabilities were most likely to be included, with an eligibility rate of 91\%.
Some AI model vulnerabilities were consistently included, whereas others were excluded relatively often.
Meanwhile, AI feature vulnerabilities were even less consistently in-scope.

\textbf{Consistently Accepted.}
Some vulnerabilities demonstrated both high mention frequency and high inclusion rates. Notably, categories under AI system like unauthorized actions, cross-customer data access, or privilege escalations are almost universally accepted (100\%). These types of vulnerabilities also exist in traditional systems. 
The vulnerabilities that are specific to AI models, such as data poisoning, model extraction, and inference attacks also showed strong eligibility rates (over 80\%), which shows that AI system and model vulnerabilities are taken seriously by vendors. 

Model extraction was consistently included, likely due to the vendor's intellectual property being at risk. Google explained that ``\textit{AI models often include sensitive intellectual property … such as their architecture or weights.}''

\textbf{Inconsistent Coverage.} The AI feature category, like harmful output and hallucination, showed more fragmented treatment. Harmful output was among the most frequently mentioned, which Priceline defined as ``\textit{harmful, inappropriate, or misleading responses}''. It had an eligibility rate of only 42\%. 

Similarly, hallucination showed moderate mention frequencies but had the lowest acceptance rate (14\%). 
However, this is likely not due to a lack of importance. Amazon explained that these issues may only be ``\textit{about responsible AI usage}'' as there is ``\textit{no clear application security impact}'' for the vendor to verify.

Although jailbreaking was mentioned 11 times, only 3 of the vendors treated these as eligible vulnerabilities, resulting in an eligibility rate of 27\%.
Atlassian and Anthropic described jailbreaking as ``\textit{multi-turn}'' and involving circumvention of the models' ``\textit{built-in security measures and ethical guidelines}.'' Google explained that the reasons for rejection include that jailbreaking only ``\textit{impacts a user's own session}'' and that ``\textit{fully mitigating [it] may be completely infeasible}.''

The main reasons for AI vulnerabilities being out-of-scope included: the ``attack'' resulting from expected model behavior, as indicated by seven vendors. Skyscanner noted that some issues are ``\textit{acceptable behavior from LLM models}''. Eight vendors mentioned insufficient demonstrable impact, for instance Asana excluded bugs with ``\textit{no measurable security impact}.'' 

Other reasons for exclusion include the lack of reproducibility (\textit{n} = 5), as Meta indicated, ``\textit{cannot consistently be reproduced}''; patching not being possible (\textit{n} = 2), for example ``\textit{cannot be fixed directly}'', as mentioned by OpenAI; and low risk (\textit{n} = 4), as highlighted by GitHub, ``\textit{... accept the low risk as a security/usability trade-off.}''

\subsection{Policy Evolution}
\label{sec:Evolution}
As shown in Figure~\ref{fig:evolution}, a total of 205 AI-related policy changes across 46 vendors were identified between 2018 and 2025. 
Among them, modifications related to scope and vulnerability eligibility accounted for the largest share, totaling more than 57\%. Specifically, \verb|Scope-In/Vuln-Eligible| changes represented 38\%, while \verb|Scope-Out/Vuln-Ineligible| accounted for 19\%.
The earliest recorded policy change involving AI vulnerabilities appeared in December 2018, when Tripadvisor explicitly excluded their AI product from the scope of the bug bounty policy. The next came in June 2020, when Asana included prompt injection in its list of eligible vulnerability categories. Updates remained infrequent in 2021 and 2022, when three vendors made AI-related adjustments, primarily involving scope additions and exclusions.

In 2023, AI-focused policy changes became more frequent, with seven companies updating their policies. Of these, five either began accepting AI vulnerability reports for the first time or expanded their disclosure frameworks to explicitly address AI products. Two others introduced or refined structured evaluation processes, including setting reward ranges and defining severity classifications specifically for AI-related findings. Four vendors also formalized exclusions for certain categories, such as model prompt extraction or off-topic chatbot responses.

At least twenty companies made AI-related updates in 2024, with thirteen explicitly adding AI vulnerabilities to their accepted scope, whether by creating new categories, clarifying applicable AI features, or establishing exclusion criteria. Nine companies refined pre-existing AI policies through expanded scope descriptions, added documentation for testers, or the introduction of new bounty pricing for specific issues such as bias or generative content misuse. Across all updates in 2024, there were fifteen instances of initial inclusion of AI vulnerabilities, eleven cases of scope or procedural refinement, eight changes to reward structures or severity grading, and seven updates defining eligible or ineligible AI vulnerability types. 

By August 2025, more than fifteen companies updated their AI vulnerability policies. Seven of these addressed AI vulnerabilities for the first time, while others focused on adjustments to reward structures, clarification of testing rules, or the introduction of compliance and participation restrictions. 

Among vendors, model providers showed the most frequent updates. Microsoft had the most AI-specific updates, primarily focusing on \verb|Reward-Evaluation|, continuously refining the evaluation criteria and pricing.
Many AI companies initially did not have a clearly defined acceptance scope for AI vulnerabilities, so their first updates often focused on \verb|Scope-In| (\textit{n} = 41). 

Even in subsequent iterations, \verb|Scope-In| remained a high-frequency change, indicating that as AI product forms and functionalities evolved, AI companies adjusted which parts could be tested and reported. In contrast, reward mechanisms (\textit{n} = 36) and vulnerability type definitions (\textit{n} = 31) also saw many changes, but these typically occurred only after scope had been established.

\section{RQ2: Profiles and Positions}
\label{sec:Profile}
Based on the granular findings in the previous section, we classify 264 AI vendors into three broad approaches.
Section~\ref{subsec:proactive} describes the approach of vendors that affirmatively clarify how AI vulnerabilities should be disclosed.
Section~\ref{subsec:silent} explains the most common approach, in which vendors do not update policies to address AI vulnerabilities.
Finally, Section~\ref{subsec:excluded} describes the vendors who explicitly exclude AI vulnerabilities or have not published a disclosure document.

\subsection{Proactive Clarification} 
\label{subsec:proactive}
The proactive profile (\textit{n} = 46) encompasses vendors that, to varying degrees, acknowledge AI vulnerabilities in their disclosure programs and take steps to encourage reports. We identify three subtypes:
a) \emph{Active Supporters}: where AI submissions are welcomed as a distinct set of security issues.
b) \emph{Integrationist}: where AI vulnerabilities are listed, but not treated differently.
c) \emph{Back-channel}: where AI issues are not in scope for rewards, but can be reported via a separate channel.

\subsubsection{Active Supporters} 
Vendors in the active supporter category (\textit{n} = 9) articulate a clear structure for the disclosure, evaluation, and remediation of AI vulnerabilities. 
These organizations demonstrate not only openness to AI security concerns, but also offer concrete technical and procedural support for external researchers. Their profiles are characterized by both clarity of scope and institutional readiness, even though they may still declare specific AI vulnerabilities as out-of-scope.
This category consists of Google, Microsoft, Meta, Anthropic, Hugging Face, Priceline, Mozilla, Atlassian, and Skyscanner, with model providers accounting for a relatively large share in this category.

\textbf{Defining \& Encouraging.} The defining feature of active supporters is explicit encouragement of AI submissions (\textit{n} = 7). This is often expressed in the \verb|Company-Statement| and \verb|Engagement| modules, for example: Anthropic ``\textit{actively seek any reports on model safety concerns... in Claude}''.
Other vendors extend the scope to unknown or novel vulnerabilities (\textit{n} = 4), as illustrated by Hugging Face's policy, which states ``\textit{These categories are not exhaustive... [we] always keep an eye out for new threat vectors.}'' 

To clarify the terrain of acceptable testing, active supporters always specify AI-specific assets to help bug hunters determine scope. The nine vendors clarify that the following products are in-scope: AI-enabled products or services (\textit{n} = 14), LLMs (\textit{n} = 5), plugins or API interfaces (\textit{n} = 3), and source code or model files (\textit{n} = 4).

Even when active supporters declare certain AI vulnerabilities as out-of-scope, they add exceptions for edge cases.
These exceptions include conditions such as demonstrable harm (\textit{n} = 5), or vulnerabilities with reproducibility or strong proof-of-concept (\textit{n} = 4). Priceline specified that AI vulnerabilities are ``\textit{reviewed on a case-by-case basis''} and will pay out on issues that can ``\textit{be exploited in a meaningful way.}''

A typical characteristic of active supporters is the articulation of clearly defined eligible AI vulnerabilities. Instead of merely listing categories, these policies provide descriptions or examples (\textit{n} = 8). Microsoft, for instance, described adversarial perturbation as ``\textit{inputs that are provided to a model that result in a deterministic, but highly unexpected output from the model.}''

Active supporters also specify scenarios for inclusion (\textit{n} = 4), such as ``\textit{only in-scope when a model's output is used to change the state of a victim's account or data}''. 
They also outline security impacts (\textit{n} = 7), such as declaring in-scope vulnerabilities that ``\textit{potentially break out of sandboxed interpreters and compromise system integrity.}'' 
For ineligible vulnerabilities, concrete reasons are likewise offered, as discussed in Section~\ref{sec:vuln}.

Three active supporters report following AI safety frameworks like the \textit{Voluntary AI Commitments}~\cite{whai23} and the \textit{Hiroshima Process International Code of Conduct for Advanced AI Systems}~\cite{hiroshima23}.
This formalizes the vendors' proactive stance toward AI security and safety. Two vendors also address the use of AI by security researchers, disclosing their use of automated testing or AI-generated reports, such as when the researcher ``\textit{used AI in the creation of the vulnerability report, you must disclose this fact.}''

\textbf{Separate Evaluation.} 
Some active supporters operate a dedicated AI vulnerability reporting channel (\textit{n} = 5), such as Google's ``\textit{Abuse Vulnerability Reward Program}'', or maintain a clearly separated track within an existing VDP/BBP framework (\textit{n} = 3). This structure enables more streamlined triage and prioritization of AI bug reports, as reflected in Google's statements like: ``\textit{...abuse risks are often inherent to product features, ... [we] have a dedicated reporting channel for these types}'' and ``\textit{Identifying and mitigating universal jailbreaks is the focus of this bug bounty program.}''

Four active supporters provide AI-specific severity rating frameworks. For example, Microsoft provided a standalone table that ``\textit{describes the severity classification for common vulnerability types in systems involving AI/ML.}''
Meanwhile, Google referenced ``\textit{impact assessment}'' and ``\textit{probability assessment}'' providing details about how these are calculated. 
Similarly, Meta clarified that ``\textit{the following factors when deducting from the payout:}'' such as ``\textit{limited internal access}'' or ``\textit{assets are intentionally scrapable.}''
 
Vendors without a dedicated framework also outline AI-specific severity–reward structures (\textit{n} = 6) separately, in order to clarify whether these are higher or lower than those for traditional vulnerabilities, such as ``\textit{with a maximum payout... specific to GenAI products.}''
In addition, three vendors' policies embed recommendation requirements directly in submission guidelines, asking researchers to propose mitigation strategies alongside their reports, for example: ``\textit{... focus on having appropriate and valid recommendations.}''

\textbf{Resources.}
Active supporters also invest in the supporting materials necessary to facilitate meaningful testing. This includes publishing AI testing guides (\textit{n} = 5), such as Atlassian's note: ``\textit{Please review the documentation here to ... test drive its features and functionalities.}''

Vendors also provide cases for testing (\textit{n} = 5), for example, Skyscanner provided an ``\textit{Example Query for Testing}'' to guide vulnerability hunters in conducting the ``\textit{AI Bias Testing.}'' Active supporters also provide educational resources (\textit{n} = 4), such as ``\textit{... a tutorial video series for complete beginners...}''; and provide links to academic references (\textit{n} = 3).
Such resources may reflect an understanding that AI risks often require context-specific knowledge or non-standard testing methodologies.

\subsubsection{AI Integrationists}
Twenty-nine AI vendors fold AI vulnerabilities into the existing disclosure policy without providing standalone encouragement, guidelines, or dedicated testing materials.
Notable vendors in this category include Amazon, Apple, Adobe, GitLab, and Nvidia.
These vendors typically clarify which AI products are in scope, or list categories of AI vulnerabilities that are either in scope or excluded. 

\textbf{Reliance on Lists.} One of the most recognizable features of integrationist vendors is reliance on checklist-style enumeration. 
This involves providing lists of the scope of AI-related products (\textit{n} = 21) or eligible AI vulnerability types (\textit{n} = 12), without offering further descriptions, reasoning, or attack context. For example, Glean listed the following AI vulnerabilities as in scope: ``\textit{LLM related attacks: Data exfiltration; Insecure output handling.}'' 
These lists are the most affirmative approach in this category.

A small subset of these vendors (\textit{n} = 4) reference AI vulnerabilities using overly broad terminology, such as ``\textit{model safety issues}'' or ``\textit{... issues related to model prompts and responses.}''
Eight vendors fail to clearly distinguish AI-related vulnerabilities from traditional security issues. 
In these cases, AI vulnerabilities are subsumed under broader categories of conventional issues, which leaves researchers to identify and interpret the relevant scope on their own. 

Integrationist vendors also provide little explanation for why certain categories are out-of-scope, offering only lists (\textit{n} = 9). These vendors only accept well-known attack types (\textit{n} = 5), with no acknowledgment of new AI vulnerabilities.
The lack of additional details or explanation may reflect the vendors' belief that AI vulnerabilities are a subset of traditional security that does not need further explanation. It is unclear whether this view is shared by security researchers, who may appreciate affirmative clarification.

\textbf{Lack of Resources}. Four integrationist vendors add links to relevant developer documentation or AI product usage pages. Only two vendors specifically offer guides tailored to AI scenarios. This lack of AI-specific guidelines and resources may hinder researchers' ability to find, test and validate AI issues.

From a structural perspective, there is no separate AI program, pricing, or evaluation mechanism. This may result in reduced signal clarity, and researchers have no way of knowing how their reports were evaluated or the range of amounts they might have been paid.

\subsubsection{Back Channel}
Back channel vendors offer an alternative channel for submission (\textit{n} = 8). Representative vendors in this category include OpenAI, Dropbox, Cohere, Canva, and GitHub. These vendors mention AI issues in their policies but do not provide a formal reporting mechanism; instead, they offer an alternative channel for submitting AI-related issues.

\textbf{Alternative Path.} These vendors typically do not treat AI issues as security vulnerabilities. For example, Brave stated that ``\textit{AI misuse}'' or ``\textit{issues with models}'' are not valid vulnerabilities and are therefore ineligible for the policy. 

However, they still offer an alternative channel outside the disclosure program, typically in the form of online forms, in-app reporting features, dedicated contact emails (e.g., safety@vendor.com), or simply by noting that AI issues should be ``\textit{reported to the AI developer team.}''
No rewards are offered for submitting to these channels.

\subsection{Silent} 
\label{subsec:silent}
Most vendors fall into the silent profile (\textit{n} = 115).
These vendors have not updated policies to address AI vulnerabilities.
This also includes silent vendors whose hosted platforms or their own programs allow AI-related reports. The well-known vendors in this category include Cursor, Huawei, Intel, Zoom, and Samsung.
Researchers may still report AI issues but without guidance or assurance.

\begin{figure}[]
\hspace*{-0.4cm}
    \centering
    \includegraphics[width=0.42\textwidth]{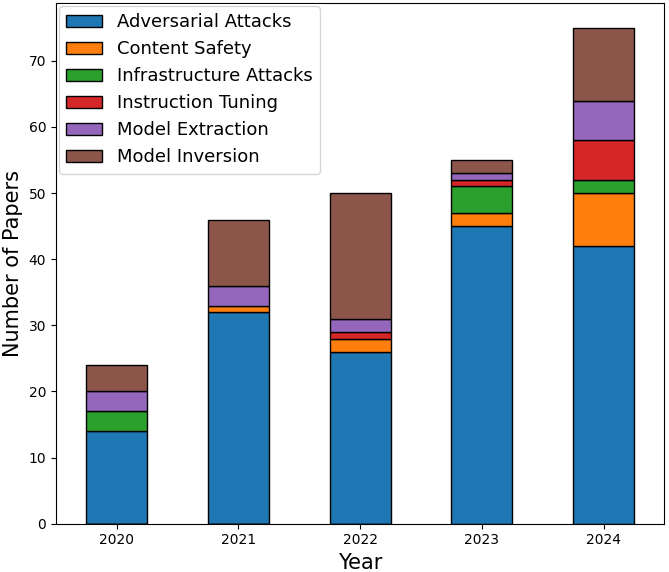}
    \caption{Yearly distribution of AI security \& safety papers by topic based on meta-taxonomy in Appendix~\ref{sec:Taxonomy}.}
    \label{fig:paper}
\end{figure}

\subsubsection{Self-Hosted Policies} 
The first type consists of vendors (\textit{n} = 66) whose policies contain no AI-related content and provide no alternative channels.
This was most common for vendors in Gen-AI App and Data Science \& ML Platform categories.
Even when policies are silent, AI vulnerabilities may be accepted when they qualify under existing terms and conditions.

For example, Cursor's policy does not mention AI vulnerabilities.
However, Cursor has patched AI vulnerabilities\footnote{\url{https://github.com/cursor/cursor/security/advisories/GHSA-vqv7-vq92-x87f}}, such as ``\textit{Arbitrary code execution from Cursor Agent through a prompt injection.}''
Even though AI vulnerabilities may be accepted in practice, this may not be clear to researchers.

\subsubsection{Hosted Policies} 
Many vendors who host their policy on platforms (\textit{n} = 49) have not updated their policies to address AI security.
However, the bug bounty platform has added a specific submission field for AI vulnerabilities.
For example, both HackerOne and Bugcrowd have incorporated the \textit{OWASP Top 10 LLM}~\cite{owasp-llm-top10} into their taxonomies of vulnerabilities. This allows hunters to select relevant vulnerability categories at the time of submission. 
This makes AI vulnerabilities reportable, but there is no certainty about whether they are in or out of scope.

\subsection{Restrictive} \label{subsec:excluded}
Restrictive AI vendors (\textit{n} = 103) can be grouped into two categories: those companies with no vulnerability disclosure channel at all; and vendors that explicitly refuse to accept any AI vulnerability reports. Examples of vendors in this category include Otter, Azwedo, Tripadvisor, Skroutz, and Nubank.

\textbf{No Policy.}
Many AI vendors (\textit{n} = 96) provide no vulnerability disclosure channel at all.
This means we did not identify a bug bounty or responsible disclosure policy, or even an email address for a security contact.
Some vendors in this category may be misclassifications because we could not find the contact channel, despite repeated checks (see Section~\ref{sec:Collection}).
However, we expect that researchers would find it similarly difficult to identify the channel, given we checked at least three sources.

\textbf{Excluded.} 
A few vendors explicitly reject AI vulnerabilities (\textit{n} = 7). 
Some vendors exclude AI products from the list of assets that are in scope, such as Skroutz, which declared ``\textit{our AI-chatbot agent}'' as out-of-scope.
Similarly, Tripadvisor stated that ``\textit{Attacks against our artificial intelligence systems are ineligible}'' in their bug list. 
However, this exclusionary posture may change with time. For example, Nubank stated that ``\textit{reports related to LLM Based Applications are temporarily out of scope,}'' which suggests that this may be a short-term measure.

\section{AI Incident and Research Alignment (RQ3)}
This section explores the extent to which AI disclosure policies align with actual incidents and also research.
This research question compares our findings to overarching societal and research trends, which are inherently difficult to quantify (discussed in Section~\ref{subsec:limitations}).
As such, the results in this section do not offer the same degree of precision or definitiveness as the results discussed earlier. However, it still provides useful insight into how industry practices relate to real-world harms and research priorities.

\subsection{Timing} 
\begin{figure}[]
\hspace*{-0.4cm}
    \centering
    \includegraphics[width=0.51\textwidth]{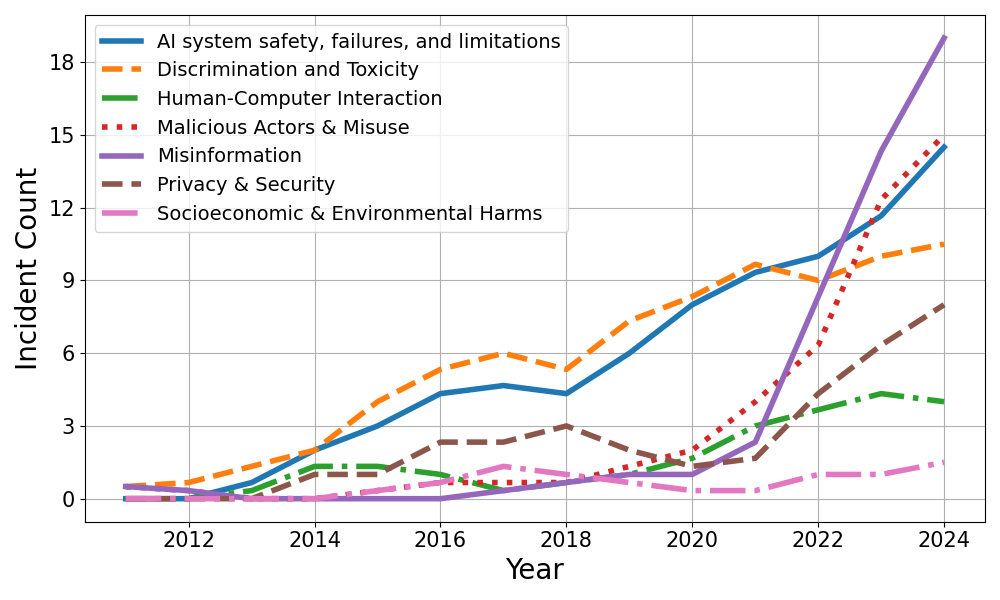}
    \caption{Monthly distribution of AI incidents impacting vendors in our corpus across MIT risk domains~\cite{Risk24}.}
    \label{fig:mit}
\end{figure}

Vendor disclosure policies affirmatively addressed AI vulnerabilities later relative to both academia and actual incidents.
The first disclosure policy in industry that mentioned AI was Tripadvisor's in late 2018, followed by Asana in 2020, Coveo in 2021, and Sendbird and Dropbox in 2022 (see Figure~\ref{fig:evolution}).
By contrast, there were public reports of AI incidents as early as 2012 (see Figure~\ref{fig:mit}).
Academics had already been asking questions like ``can machine learning be secure?'' since 2006~\cite{barreno2006can}. A stream of seminal papers suggested the answer was ``no''~\cite{dalvi2004adversarial, szegedy2013intriguing, goodfellow2014explaining, carlini2017towards}, which motivated disclosure policies. Focusing on a specific AI vulnerability, the first academic model extraction attack was published in 2016~\cite{tramer2016stealing}. As shown in Figure~\ref{fig:paper}, model extraction and inversion accounted for a quarter of security articles throughout 2020–2024. Yet, the first vendors to affirmatively clarify that such vulnerabilities were in scope were Google and Microsoft in 2023. 

However, the delay was much shorter in the case of vulnerabilities associated with LLMs, such as jailbreaking and prompt injection. 
Both types of AI vulnerability were widely discussed on social media platforms following the release of ChatGPT in late 2022 and into 2023~\cite{wei2023jailbroken, AAAI21}. 
Meanwhile, vendors added an LLM vulnerability to their disclosure policy as early as March 2020, when Asana published a ``\textit{Special note on prompt injection}'' stating that it would ``\textit{consider submissions of this type.}''
By contrast, academics had studied how adversarial tokens could manipulate model outputs as early as 2019~\cite{wallace2019universal}. In the past four years, adversarial attacks and instruction tuning together accounted for more than 64\% of AI security papers.
However, the delay addressing prompt injection was much shorter, relative to model stealing~\cite{tramer2016stealing} and model extraction.
The research may have been less directly analogous to the AI vulnerabilities that were added to disclosure policies.
Regardless, all three datasets have experienced hockey-stick growth in the last few years.
This growth was sharpest for disclosure policies.
The AI incidents and academic articles in 2023 and 2024 accounted for, respectively, 45\% and 46\% of the total dataset.
The equivalent proportion for AI-mentions in disclosure policies was 70\%.

\subsection{Types of Vulnerabilities} 
Across the 319 incidents impacting the vendors in our sample (2011--2024), Figure~\ref{fig:mit} shows that discrimination and toxicity (26\%) and AI system safety, failures, and limitations (25\%) together accounted for just over half of all incidents. 
Misinformation accounted for 15\%, of which the vast majority (94\%) involved false or misleading information. 

By contrast, AI disclosure policies focused on technical categories such as system integrity, authentication/authorization, model extraction, data leakage, and prompt injection (see Table~\ref{tab:vuln-types}). These map directly to categories in the incident taxonomy like \textit{AI system security vulnerabilities and attacks} and \textit{lack of capability or robustness}, which together accounted for 29\% of all incidents. 
Similarly, data leakage or information disclosure vulnerabilities accepted in policies directly correspond to \textit{compromise of privacy} incidents, which account for 11\% of the dataset.

Although incident counts are not company-specific, the distribution of mentions highlighted how risks are concentrated around particular vendors. OpenAI accounted for roughly one-fifth of all company references, and it dominated discussions of misinformation (26\%). Notably, OpenAI considered such issues as ``\textit{strictly out of scope},'' since ``\textit{they are not standalone vulnerabilities...}'' and ``\textit{Solving these issues usually requires significant research and a broader approach.}''
Meanwhile, Google appeared most often in discrimination cases (23\%), Meta was the leading actor in safety-related mentions (25\%), while Twitter/X dominated the misuse category (28\%).

There is potentially some misalignment. Vendors like Amazon, Meta, and Microsoft dedicated significant attention to model theft and adversarial examples, yet these subdomains accounted for less than 10\% of publicly reported incidents. 
Meanwhile, public reporting on AI incidents was dominated by downstream harms such as \textit{toxic content exposure} and \textit{false or misleading information} (47\% across both categories). 

Only a minority of companies explicitly included content safety or harmful output as reportable vulnerabilities.
Much like vendor disclosure policies, academic computer security research has focused primarily on model-level vulnerabilities such as adversarial examples, poisoning, and model extraction. These categories together accounted for over half of the AI security \& safety paper dataset.

\section{Discussion}
\label{sec:discussion}
We discuss the implications for academic debates around what constitutes AI security, recommendations regarding what industry should do, and the limitations of our study.

\subsection{The Nature of AI Vulnerabilities}
The disclosure practices vary across vendors, which belies different perspectives on AI security.

\textbf{AI Vulnerabilities Fit into Existing Disclosure Policies.} 
When vendors simply add AI products to a list of products that are in-scope for rewards, they implicitly assume the existing policy covers AI vulnerabilities, provided they meet the criteria for severity.
This makes sense given vendors aim to maintain security properties.
If a vulnerability leaks sensitive user data, the vendor should pay to fix it regardless of whether the bug is a SQL or prompt injection.

This view is supported by Table~\ref{tab:vuln-types}, which shows that AI vulnerabilities are most likely to be in-scope when described in terms of which security policies are violated (e.g.\,system integrity, data access, or authorization).
By contrast, AI vulnerabilities are more likely to be excluded when defined in terms of buzzwords (e.g.\, jailbreak) or downstream consequences outside traditional security models (e.g.\, hallucination or harmful output).
In this interpretation, the new aspect of AI is the increased expectations regarding harm prevention.
For comparison, spell-check vendors are not expected to stop their product improving spelling and grammar in abusive and factually incorrect content. 
Most vendors appear to take this view, given 82\% of vendors do not have specific AI disclosure policies.
However, some of them may update policies to address AI security in the future.
After all, vendors who actively support AI disclosures are also the vendors that process the most vulnerabilities (e.g., Microsoft and Google) and BBPs like HackerOne and BugCrowd.
The collective experience of them motivates consideration of an alternative view, which the industry may or may not be in the process of adopting.

\textbf{AI Vulnerabilities Require Updated Policies.}
Evidence for the uniqueness of AI security can be seen in policies that direct submissions to an alternative submission channel, often an email address like ``\textit{ai-safety@vendor.com}.''
Future work should examine which department receives these submissions.
A provocative simplification might be that software engineers receive traditional bug reports, whereas data scientists receive AI bugs.
This raises the question of how a data scientist can ``patch'' an AI vulnerability by, for example, re-training the model. OpenAI argued: 
``\textit{Model security issues are not well suited for bug bounty programs because they are not standalone vulnerabilities that can be fixed directly.}''
Google's Abuse VRP argued similarly with the statement: ``\textit{abuse risks are often inherent to product}''.
These statements suggest AI vendors are shipping products that they acknowledge can never be fixed, or at least not directly. 

More optimistically, safeguards will improve with time and attention, a view that motivated Anthropic to expand their bug bounty to exactly this.
This approach sees academic and industry research focus on improving defenses that compensate for the fundamental insecurity of AI models.
However, this sounds familiar to what happened with traditional security, with Ross Anderson characterizing the strategy as: ``ship it Tuesday and get it right by version three''~\cite{anderson2010security}.
The AI industry may follow the software industry in not building products that are secure by design.
The parallels are uncanny in that foundational research finds modern computers~\cite{thompson1984reflections, anderson2010security, schneier2015secrets}, neural networks~\cite{goodfellow2014explaining, tramer2020adaptive}, and LLMs~\cite{wei2023jailbroken, bommasani2022opportunitiesrisksfoundationmodels} are too complex to be made 100\% secure.
Instead defenders will have to buy bolt-on security products like LLM filters.
Both industry and academia benefit from the cycle of proposing defenses, finding attacks that break defenses, and repeating it~\cite{herley2017sok}.

\subsection{Recommendations}
\textbf{Broaden Disclosure Policies.}
AI vendors with no disclosure channel (36\%) may be early-stage startups that have not yet built out formal security processes, as vulnerability maturity develops with experience~\cite{sridhar2021cybersecurity}. Such firms should at least publish an email contact for bug reports. 
The vendors that operate a VDP (16\%) without offering rewards should consider financially compensating researchers for their time.
It is less clear whether the 45\% of AI vendors that do not mention AI vulnerabilities should clarify whether such vulnerabilities are in or out of scope. Answering this requires a qualitative study involving AI vendors.

\textbf{Clarify Supply Chain Reporting Channels.}
Many AI applications rely on foundation models, which means AI vulnerabilities in their products may actually affect models that application vendors do not control.
This may explain why more model providers maintain AI disclosure programs (48\%) than do other AI vendors (17\%).
Researchers may not know whom a vulnerability report should be sent to, especially if the foundation model dependency is not transparent.
AI application vendors need to either run a disclosure program and forward relevant reports to the model provider, or clearly disclose which AI services they depend on and point researchers towards their appropriate disclosure policy.

\textbf{Monitor Safety Risks.}
Content safety issues like discrimination, toxicity, and misinformation account for 41\% of publicly reported AI incidents (see Figure~\ref{fig:mit}). Yet just 5\% of vendors consider these kinds of issues as in-scope for vulnerability reporting.
Content safety is declared out-of-scope via reasoning like ``abuse risks are often inherent to [AI] product features''.
Rather than argue these issues should be in-scope, we recommend that vendors at least track and monitor safety-related issue trends via dedicated safety reporting mechanisms and usage monitoring.
OpenAI's malicious use reports are a step in the right direction\footnote{\url{https://openai.com/global-affairs/disrupting-malicious-uses-of-ai-june-2025/}}, but data sharing should allow external researchers to empirically study the harm and how to calibrate the threshold.

\textbf{Future Work.}
Future work should revisit our research questions once adoption among AI vendors increases in the long tail. Researchers should conduct studies with bug hunters to understand their experiences and challenges with current AI vulnerability disclosure practices. Additionally, comparative studies could examine how different disclosure policy designs affect the real-world reporting and remediation of AI vulnerabilities.

\subsection{Limitations} 
\label{subsec:limitations}
We acknowledge several limitations, and present the main potential sources of bias that may affect our results.

\textbf{Focus on Affirmative Discussion.}
Our analysis focused on vendors that explicitly mention AI-related vulnerabilities.
Studying affirmative policies provides insight into strategies that actively address vulnerabilities in AI systems.
However, this does not mean the other vendors do not accept AI vulnerabilities, as they may be accepted under generic terms and conditions not written with AI in mind.

\textbf{Sampling AI Vendors.}
Our definition of an AI vendor followed Gartner's product classifications, which ensured that our results apply to the most significant AI technologies like foundation models, ML platforms, and Gen-AI apps.
Our sample included the largest vendors (OpenAI, Google, Meta, Apple, DeepSeek, Huawei), making it representative of mainstream AI products used by millions of users. However, our results may not generalize to niche or AI-lite products offered by smaller vendors using off-the-shelf AI components in less sophisticated applications.
Preliminary analysis suggests these kinds of vendors have not adopted affirmative AI vulnerability disclosure yet. Tracking this is an important area for future work.

\textbf{Public AI Incidents.}
The AI Incident Database~\cite{AAAI21} has a bias toward incidents that are more visible or newsworthy, as it relies heavily on media reports and public submissions.
This may, for example, over-represent misinformation incidents as they aim to generate attention, which facilitates detection, and the political salience likely drives news coverage.
By contrast, technical security vulnerabilities may not receive timely public reporting, either because they are quietly fixed or because the attacker remains undetected.

\textbf{Technical Literature.}
It is simply infeasible to conduct a comprehensive literature review of ``vulnerabilities'' in AI systems.
Sampling from four security conferences biases our sample towards technical security issues, which aligns with the kind of issues that bug bounty programs have covered historically~\cite{Walshe23}.
However, this focus may omit representative work published in other AI venues which also produce influential research on AI security.
A different approach could have focused on human-centered research into how AI enables discrimination~\cite{birhane2022forgotten}, perceptions of unfairness~\cite{binns2018s}, psychological harm~\cite{chandra2025lived}, and other emerging topics.

\textbf{Private AI Security Measures.}
The ``lag'' between incidents exploiting AI vulnerabilities and these being added to disclosure policies is not necessarily an oversight by vendors.
AI vulnerabilities may have been accepted under existing policies, especially when vulnerabilities impact integrity and confidentiality.
Vendors may also address these issues via internal testing, guard rails, and other internal security processes that our methodology cannot observe.
In addition, we did not assess other AI- or security-related documents or public commitments, did not examine red teaming, and cannot measure private bug bounty programs.
Thus, our findings did not reflect the full extent of vendors' AI security practices.

\section{Conclusion}
Our study examines vulnerability disclosure policies in the AI industry across 264
vendors. 
Only 64\% of AI companies maintain a disclosure channel, and explicit recognition of AI vulnerabilities is limited (18\%), with approaches varying widely. 
Model providers were most likely to explicitly include AI vulnerabilities; meanwhile many AI application providers had generic documents that did not address AI security.

AI system vulnerabilities are most consistently declared in-scope (91\%), and issues related to AI features are most commonly declared out-of-scope (50\%). With respect to AI companies' positions on AI vulnerabilities, we identified three profiles: proactive clarification, silent, and restrictive. 
Vendors were later to update disclosure documents to address AI risk relative to when these issues were identified in academic papers and real-world incidents.
Disclosure policies and academia have more focus on upstream technical issues in isolation, relative to the AI incidents that are dominated by content safety issues.


\appendix
\section*{Ethical Considerations}
We evaluated the ethical implications of our research and consider the associated risks to be relatively low given our study involved no human subjects or data about individuals. Below we outline the main ethical considerations. 

\noindent
\textbf{Stakeholders.}
The primary stakeholders of this research include: the AI companies whose disclosure policies were analyzed (i.e., AI vendors); security researchers and bug hunters who may use or be affected by these policies; end users who are indirectly impacted by the security posture of AI vendors; policymakers; and the broader public. The research team itself is also a stakeholder, responsible for data management.

\noindent
\textbf{Potential Impacts on Stakeholders.}
For AI vendors, we analyze publicly available policy documents and may highlight aspects of their disclosure mechanisms, which could impact their reputation (we expand on this in the Decision part).
For security researchers, our work helps to identify ambiguous policy areas and offers guidance that may support AI vulnerability submissions.
For end users and the broader public, our findings may encourage more transparent and responsible AI vulnerability disclosure, ultimately improving AI product security.
For policymakers, the study provides empirical insights into current industry practices, supporting future standards and regulatory frameworks.

\noindent
\textbf{Data Collection.}
All vulnerability reporting policy data and incident data analyzed were obtained from company websites or databases that are public online. 
We focused on documents related to company policy, which contained no personal information related to individuals.
Our data acquisition process was designed to respect the integrity and availability of the source websites: 1) we accessed each website only a limited number of times; 2) our automated data-gathering scripts did not log in or remain on the websites; and 3) operations were conducted at a carefully controlled rate to avoid imposing undue load on servers or disrupting normal operations.
We respected the terms of service of websites, reverting to manual search if required by the websites.

\noindent
\textbf{Disclosure.}
We shared the dataset publicly to support future study. Our work focuses on policy analysis; hence we do not discover any new vulnerabilities or interact with live systems for exploitation.
We shared the real names of companies to improve transparency and accountability.
By explicitly naming companies, we hope that our results will help vendors reflect on how disclosure policies are designed and communicated, as well as rethinking the nature of AI vulnerabilities. 

\noindent
\textbf{Conflict of Interest.} 
The authors have no conflicts of interest with any of the companies analyzed. 

\noindent
\textbf{Decision.} We identified two main risks: 1) our results might negatively impact the reputation of the vendors in our sample; and 2) we would face repercussions as a result.
These risks need to be traded off against the potential to increase transparency and accountability in the AI industry, which brings benefits for directly affected end users.
To reduce both risks, we relied on direct quotes when mentioning specific vendors.
We judged that the residual risk was tolerable for the researchers, and was justified by the stated policies of vendors.
For these reasons, we proceeded with and published this study.

\section*{Open Science}
To promote transparency, reproducibility, and open scientific practice, we shared our research artifacts. In particular, we created a publicly accessible repository that includes the list of AI vendors, complete codebook for Appendix~\ref{sec:codebook}, disclosure policy dataset, coding results, longitudinal data collection script, and supplemental files for Appendix~\ref{sec:Supplement},~\ref{sec:Taxonomy} and~\ref{sec:MIT}. We make these artifacts available at \url{https://doi.org/10.5281/zenodo.17873559}. This ensures that our work can be replicated and verified by others in the community.

\bibliographystyle{plain}
\bibliography{bib.bib}

\section{Appendix}

\subsection{Methodological Supplement}
\label{sec:Supplement}
\textbf{AI-Related Categories.} For the AI-related categories used by Gartner~\cite{GartnerPeerInsights}, we selected nine that represent the most important segments of the AI ecosystem: \textit{Generative AI Apps}, \textit{AI Code Assistants}, \textit{Generative AI Engineering}, \textit{Generative AI Model Providers}, \textit{Conversational AI Platforms}, \textit{AI Applications in IT Service}, \textit{AI-Augmented Software Testing Tools}, \textit{Data Science \& ML Platforms}, and \textit{AI \& Data Analytics Service Providers}. We did not include categories like \textit{Generative AI Consulting} that had little influence on product security. We believe the results are representative of the mainstream AI products, given our sample includes the largest vendors.

\textbf{Disclosure Policy Structure.}
Walshe \textit{et al.}~\cite{Walshe23} define a general vulnerability disclosure policy framework consisting of 12 elements. These include \texttt{Company-Statement}, \texttt{Scope-In}/\texttt{Scope-Out}, \texttt{Vuln-Eligible}/\texttt{Vuln-Ineligible}, \texttt{Guideline-Submission}, \texttt{Engagement}, \texttt{Legal-Clauses}, \texttt{Prohibited-Action}, \texttt{Reward-Evaluation}, \texttt{Guideline-} \texttt{Disclosure}, and \texttt{Participant-Restriction}. The brief descriptions of these elements are provided in Table 1 (in Supplemental File\footnote{\url{https://doi.org/10.5281/zenodo.17873559}\label{fn:original}}).

\textbf{Keyword Selection.} For keyword searches, we relied on glossaries from sources including NIST's Adversarial ML Taxonomy\footnote{\url{https://doi.org/10.6028/NIST.AI.100-2e2025}}, OWASP's GenAI Security Project Glossary\footnote{\url{https://genai.owasp.org/glossary/}}, and SIA's Glossary of AI Terms\footnote{\url{https://www.securityindustry.org/wp-content/uploads/2025/02/SIA-Glossary-of-AI-Terms.pdf}}.

\subsection{Codebook}
\label{sec:codebook}

\begin{itemize}[leftmargin=1.5em, itemsep=0.5em]
  \item \textbf{AI Vulnerability Eligibility \& Scope} 
    \begin{itemize}[leftmargin=1.5em, itemsep=0.3em]
      \item Listed Eligible AI Vulnerability 
      \item Recognized Vulnerability Definitions	
      \item Acceptance Rationale for Vulnerability 
      \item Encouragement of AI-related Submissions 
      \item Exceptions to Ineligible AI Vulnerabilities 
      \item List of AI-Specific Assets/Scope 
     \item  Eligible Vulnerability Cases
    \end{itemize}
  \item \textbf{Ineligible AI Vulnerability} 
    \begin{itemize}[leftmargin=1.5em, itemsep=0.3em]
      \item Excluded Vulnerability Categories 
      \item Excluded Vulnerability Definitions 
      \item Exclusion Examples 
      \item Justifications for Exclusion 
      \item Out-of-Scope AI Products 
    \end{itemize}
  \item \textbf{Assessment \& Evaluation Criteria} 
    \begin{itemize}[leftmargin=1.5em, itemsep=0.3em]
      \item Separate Severity Rating 
      \item Focus on Recommendations 
      \item Exceptions to Standard Evaluation Criteria 
      \item Separate Pricing 
      \item AI-specific Evaluation Justifications 
    \end{itemize}
  \item \textbf{Testing \& Supporting Materials} 
    \begin{itemize}[leftmargin=1.5em, itemsep=0.3em]
      \item Testing Guide 
      \item Test Cases or Sample Inputs 
      \item Support Materials or Tools 
      \item References or Background Resources 
    \end{itemize}
  \item \textbf{Program Structure \& Submission Process} 
    \begin{itemize}[leftmargin=1.5em, itemsep=0.3em]
      \item Separate Program or Track 
      \item Alternative Submission Channel 
      \item Submission Guidelines 
      \item Clarification of Prohibited Conduct 
    \end{itemize}
\item \textbf{Targets} 
      \begin{itemize}[leftmargin=1.5em, itemsep=0.3em]
      \item \textbf {AI System} 
    \begin{itemize}[leftmargin=1.5em, itemsep=0.3em]
    \item System Integrity 
    \item Supply Chain 
    \item Resource Consumption 
    \item Data Access 
    \item Authorization 
      \end{itemize}
      \item \textbf {AI Model} 
          \begin{itemize}[leftmargin=1.5em, itemsep=0.3em]
    \item Prompt Injection 
    \item Model Extraction 
    \item Jailbreaking 
    \item Inference 
    \item Data Poisoning 
    \item Adversarial Example 
    \item Data Leakage
      \end{itemize}
      \item \textbf {AI Feature} 
          \begin{itemize}[leftmargin=1.5em, itemsep=0.3em]
    \item Policy Violations 
    \item Information Disclosure 
    \item Harmful/Insecure Output 
    \item Hallucination 
    \item Others
      \end{itemize}
    \end{itemize}
  \item \textbf{MISC}

\end{itemize}

\subsection{Meta-Taxonomy}
\label{sec:Taxonomy}
\textbf{Literature Review.} Our goal is to collect papers on taxonomies and reviews concerning AI attacks and vulnerabilities, and to construct a meta-taxonomy using a deductive method. First, we specified the time range, data sources, and keywords. We limited our selection to peer-reviewed journal articles and conference papers. We conducted searches in three academic databases (i.e. IEEE Xplore, ACM Digital Library, ScienceDirect) with keywords related to the taxonomy (such as survey, SoK, taxonomy, etc.), AI (machine learning, large language models, generative AI, etc.), and security \& safety (attack, vulnerability, privacy, etc.).

We restricted our search to papers published after 2020. This decision was made to align the literature collection period with the time span covered by the security conference dataset used later for verification. We identified a total of 458 articles and manually reviewed titles and abstracts for relevance to our theme, resulting in 146 candidate papers during the initial screening phase. 

For candidate papers, we established inclusion criteria. We selected only full-length papers that focus on offensive security rather than defensive measures, do not solely address narrow sub-domains (e.g., backdoor attacks only), and provide a high-level classification. The mere mention in the introduction or related work sections was insufficient for inclusion. 
One of the authors read the full paper and made a decision about its relevance to our study. 
After analyzing 11 papers and checking two additional papers without identifying new taxonomy items, the meta-taxonomy was jointly reviewed and validated through discussion among three researchers.

\textbf{Meta-taxonomy Verification.} To ensure the comprehensiveness of our meta-taxonomy, we utilized a dataset of AI security papers published at the four security conferences between 2020 and 2024 for verification (IEEE S\&P, USENIX Sec, NDSS and CCS)\footref{fn:big4}. This dataset represents research focusing on technical contributions, unlike the survey papers used to construct the taxonomy. Through manual review of 359 papers, we excluded studies focused on AI-based detection methods, AI-driven vulnerability discovery, and usability-related topics. This resulted in a set of 260 papers. We then supplemented this set with any emerging or previously unobserved vulnerability types. 
Finally, as shown in Table 2 (in Supplemental File\footref{fn:original}), we developed a meta-taxonomy comprising six categories of AI vulnerabilities.

\subsection{MIT AI Risk Repository}
\label{sec:MIT}
Table 3 (in Supplemental File\footref{fn:original}) shows the MIT risk categories\cite{Risk24} used in the AI incident database and the mapping between them and the AI vulnerability types identified in the policy elements. 
To map the vulnerability themes found in vendor disclosure policies with the MIT AI risk categories\footnote{\url{https://airisk.mit.edu/}}, we adopted a semantic and functional mapping strategy. The mapping process was as follows: sub-themes were matched to risk categories based on semantic similarity between the policy descriptions and taxonomy definitions. In cases of ambiguity, we privileged mappings that best reflected the most direct and demonstrable threat surface, while allowing multiple mappings.

Most of our AI vulnerabilities were mapped to the MIT category \textit{2. Privacy \& Security}.
The sub-theme with the most mapped vulnerabilities was \textit{2.2. AI system security vulnerabilities}.
This makes sense given our seed sample was dominated by systems security conferences.
We additionally mapped data leakage and inference to \textit{2.1. Compromise of privacy}.
Within the AI features, policy violations, harmful output, hallucination, and content safety were mapped to \textit{1.2. Exposure to Toxic Content} and \textit{3.1. False or misleading content} depending on whether they represent toxic or misleading outputs, while information disclosure was mapped to \textit{2.1. Compromise of privacy}. This mapping situates each sub-theme within the MIT framework.

\end{document}